\documentclass[fleqn,10pt]{wlscirep}
\usepackage[utf8]{inputenc}
\usepackage{siunitx}
\usepackage[T1]{fontenc}
\title{Influence of Ferromagnetic Interlayer Exchange Coupling on Current-induced Magnetization Switching and Dzyaloshinskii-Moriya Interaction in Co/Pt/Co Multilayer System}

\author[1,2,*]{Krzysztof Grochot}
\author[3,**]{Piotr Ogrodnik}
\author[1]{Jakub Mojsiejuk}
\author[4]{Piotr Mazalski}
\author[4]{Urszula Guzowska}
\author[1]{Witold Skowroński}
\author[1,2]{Tomasz Stobiecki}
\affil[1]{Institute of Electronics, AGH University of Science and Technology}
\affil[2]{Faculty of Physics and Computer Science, AGH University of Science and Technology}
\affil[3]{Faculty of Physics, Warsaw University of Technology, Warsaw, Poland}
\affil[4]{Faculty of Physics, University of Bialystok, Bialystok, Poland}

\affil[*]{grochot@agh.edu.pl}
\affil[**]{piotr.ogrodnik@pw.edu.pl}

\begin{abstract}
This paper investigates the relationship among interlayer exchange coupling (IEC), Dzyaloshinskii-Moriya interaction (DMI), and multilevel magnetization switching within a Co/Pt/Co heterostructure, where varying Pt thicknesses enable control over the coupling strength. Employing Brillouin Light Scattering to quantify the effective DMI, we explore its potential role in magnetization dynamics and multilevel magnetization switching. Experimental findings show four distinct resistance states under an external magnetic field and spin Hall effect related spin current. We explain this phenomenon based on the asymmetry between Pt/Co and Co/Pt interfaces and the interlayer coupling, which, in turn, influences DMI and subsequently impacts the magnetization dynamics. Numerical simulations, including macrospin, 1D domain wall, and simple spin wave models, further support the experimental observations of multilevel switching and help uncover the underlying mechanisms. Our proposed explanation, supported by magnetic domain observation using polar-magnetooptical Kerr microscopy, offers insights into both the spatial distribution of magnetization and its dynamics for different IECs, thereby shedding light on its interplay with DMI, which may lead to potential applications in storage devices.
\end{abstract}
\begin{document}

\flushbottom
\maketitle
\thispagestyle{empty}

\section*{Introduction}

Among magnetization switching methods, spin-orbit torque current-induced magnetization switching (SOT-CIMS) in metallic multilayers and magnetic tunnel junctions offers a very short switching time (less than 1 ns) with no breakdown risk of the tunnel barrier used in typical spin-transfer torque (STT)-based memory cells. \cite{miron_perpendicular_2011, brataas_spinorbit_2014, wang_spinorbit_2017, dieny_opportunities_2020, garello_ultrafast_2014} Until recently, SOT-CIMS has been observed in a variety of heavy metal (HM)-based layered systems, such as simple HM/ferromagnet (FM) bilayers \cite{miron_perpendicular_2011,yu_switching_2014, liu_current-induced_2012}, or HM/FM/antiferromagnet (AFM) \cite{grochot_current-induced_2021,maat_perpendicular_2001,oh_field-free_2016,fukami_magnetization_2016,van_den_brink_field-free_2016} and FM/HM/FM trilayers. \cite{sheng_current-induced_2018, avci_multi-state_2017, gospodaric_multistate_2021, li_field-free_2021, lim_spin-orbit_2020, wang_time-resolved_2021, yang_multistate_2020, yun_tailoring_2020, zhang_spin-orbit-torque-driven_2019, zhu_spinorbit_2020} Heavy metal is a source of spin current due to its strong spin-orbit interactions, which cause spin current generation when a charge current flows, as claimed in Refs.\citen{dyakonov_1971,hirsh_1999, zhang_spin_2000, sinova_spin_2015}. Spin current may be injected into the FM layer as a result of the spin accumulation gradient at the interface and exerts effective SOT fields, field-like ($H_{\mathrm{FL}}$) and damping-like ($H_{\mathrm{DL}}$), which can flip the magnetization of the FM layer. \cite{emori_large_2014, yang_spin-orbit_2016, lazarski_field-free_2019, hals_phenomenology_2013, liu_spin-torque_2011, zhang_spin-orbit_2015, lau_spinorbit_2016} In systems where the FM layer is characterized by perpendicular magnetic anisotropy, the magnetization dynamics is also driven, in addition to the SHE, by a Rashba-Edelstein effect originating from the inversion symmetry breaking at the HM/FM interfaces. \cite{sinova_spin_2015, edelstein_spin_1990, cao_deterministic_2020, cui_field-free_2019, fan_quantifying_2014, ogrodnik_study_2021} To the diversity of magnetic effects present in FM/HM multilayers, one should add the interface-dependent Dzyaloshinskii-Moriya interaction (DMI), which affects both the dynamics and switching of the magnetization, as well as the domain structure at the remanent state. \cite{fertroaddmi,D0NR08594D,Szulc}

In a bilayer HM/FM system,  the magnetization state can be determined using anisotropic or spin-Hall magnetoresistance, and in a binary system two resistance stated are written using CIMS \cite{Kurenkov_DuttaGupta_Zhang_Fukami_Horio_Ohno_2019}. The spin current generated in the HM accumulates at both HM interfaces; however, it can only act on a single FM layer, potentially causing the reversal of its magnetization. In contrast, in trilayer FM/HM/FM systems, the spin current has a different polarization at both interfaces. The energy efficiency of the magnetization reversal in such trilayer systems may be slightly higher than that of the bilayer ones. Another advantage of trilayers is that they provide the possibility of four stable resistance states, making them attractive for potential use in low-power consumption and high-density memory design \cite{Lavrijsen_Lee_Fernández-Pacheco_Petit_Mansell_Cowburn_2013} and bioinspired neuromorphic computations. \cite{li_field-free_2021, lan_gradient_2021}

In this work, we present a detailed study of multilevel switching via SOT-CIMS in the Co/Pt/Co system with different Pt thicknesses. As shown in our previous work \cite{ogrodnik_study_2021}, the thickness of Pt varies along the wedge shape of the sample, resulting in a different efficiency of the spin current generation, interlayer exchange coupling (IEC) and the effective magnetic anisotropy of the two layers. For a thin Pt layer between $1$ and $2$ nm, the transition of effective anisotropy was observed from in-plane to perpendicular. We also showed a significant difference in the atomic structure of the lower Co/Pt and upper Pt/Co interfaces, which consequently affects the amount of spin current flowing into both Co layers and thus, the switching mechanisms of their magnetizations. We also discuss the role of co-existing IEC and DMI in multilevel magnetization switching. The analysis is supported by a number of experimental techniques, such as polar-magnetooptical Kerr microscopy (p-MOKE) for domain observations and Brillouin Light Scattering (BLS) to quantitatively estimate the strength of DMI. We show that the IEC can influence DMI through modifying the magnetization distribution at HM/FM interfaces. To gain insight into the physical mechanisms of the switching and enhance our understanding, we employ macrospin modeling as well as simple models of domain-wall (DW) and spin-wave (SW) dynamics. Additionally, we provide a qualitative explanation for the magnetization switching mechanism in the investigated trilayers. Finally, we demonstrate the dependence of the critical switching current on the magnitude of the IEC in the presence of DMI.

\section*{Results and Discussion}
\subsection*{Dzyaloshinskii-Moriya Interaction}
\label{sec:DMI}
In FM/HM/FM system, apart from IEC, which has been studied before \cite{ogrodnik_study_2021, lazarski_field-free_2019}, the DMI plays an important role. Pt with its substantial spin-orbit coupling, not only induces DMI but also modulates the strength of ferromagnetic IEC as its thickness varies. As both of these components are present, it is crucial to understand their contributions and relationship within the context of switching mechanism. To study the influence of the IEC on the magnitude of the DMI, we carried out BLS measurements in the Damon-Eshbach (DE) configuration on the continuous layer at several points along the Pt wedge. We showed in our previous paper\cite{ogrodnik_study_2021} that the ferromagnetic IEC decreases inversely with Pt thickness.

\begin{figure}[ht!]
\centering
\includegraphics[width=1\textwidth]{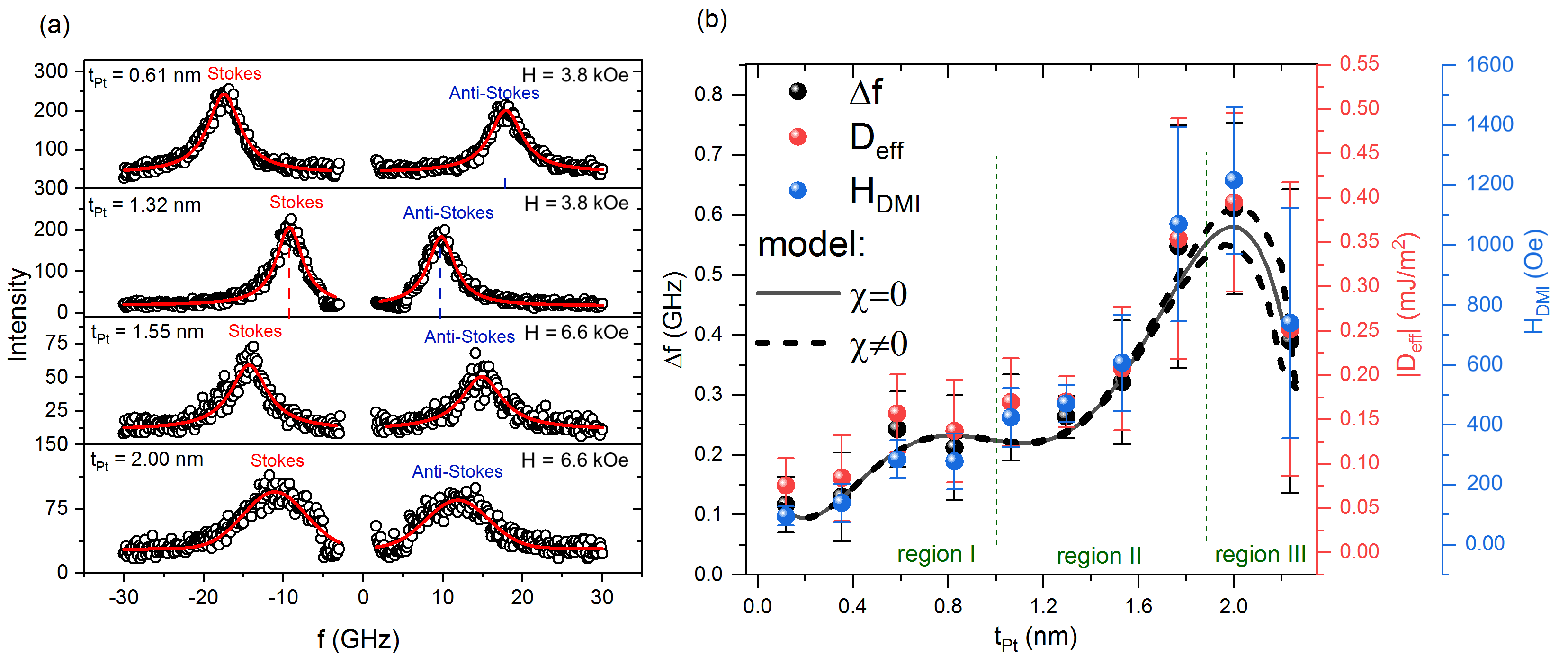}
\caption{BLS measurement to quantify DMI. (a) BLS spectra of Stokes and anti‐Stokes peaks for different Pt thicknesses from regions I-III, where the applied magnetic fields are 3.8 kOe or 6.6 kOe. (b) Extracted values of the $\Delta f$, $D_{\mathrm{eff}}$, and $H_{\mathrm{DMI}}$ as a function of Pt spacer thickness. The solid line represents the fitted theoretical $\Delta f$ calculated using experimentally derived DMI constants, where the DMI fields at the top and bottom Co/Pt interfaces are equal ($\chi = 0 $). Dashed lines: asymmetric DMI fields ($\chi \neq 0$). Upper (lower) dashed line corresponds to a 15\% higher DMI field at the top (bottom) Co/Pt interface while maintaining their average at the value used for the solid line (see Spin Wave model in Methods section for details).}
\label{fig:figure_dmi}
\end{figure}

Fig.\ref{fig:figure_dmi}(a) shows example spectra along with the Gaussian fit for several selected Pt thicknesses from regions I to III (see Fig.S1 in Supplementary Materials for the BLS spectra from a wider range of HM thickness, spectra in region IV were not measured due to equipment limitations). Larger Pt thicknesses exhibit lower intensity peaks and broader peak widths. This correlates with a significant increase in damping at the border between regions II and III and an increase in anisotropy (see Fig.S2 in the Supplementary Materials). 
For larger Pt thicknesses, it was necessary to apply a stronger in-plane magnetic field, because of the shifting of peaks toward lower frequencies when the Pt thickness increases. The frequency difference ($\Delta f$) was obtained by comparing the frequencies of the Stokes peaks ($f_{\mathrm{s}}$) anti-Stokes peaks ($f_{\mathrm{as}}$). In Fig.\ref{fig:figure_dmi}(b), the relationship $\Delta f(t_{\mathrm{Pt}})$ is shown by black dots, and the uncertainty, indicated by bars, arises due to the uncertainty of the function fitting to the BLS spectra. 

Based on the obtained $\Delta f$ values, the effective DMI constant ($D_\mathrm{eff}$) was calculated as \cite{emori_current-driven_2013}: 

\begin{equation}
D_\mathrm{eff} = \frac{\Delta f \pi M_{s}}{2 \gamma k} 
\end{equation}
where: $M_s$ is the saturation magnetization determined in our previous work \cite{ogrodnik_study_2021}, $\gamma$ is the gyromagnetic ratio of $1.76 \times 10^{11}~T^{-1} s^{-1}$ and k = 11.81 $\mu m^{-1}$ is the wave vector.

Knowing $D_\mathrm{eff}$ for each studied Pt thickness, it was possible to calculate the field $H_\mathrm{DMI}$, according to the formula \cite{emori_current-driven_2013}:
\begin{equation}
H_\mathrm{DMI} = \frac{D_\mathrm{eff}}{\mu_0 M_s \Delta} 
\end{equation}
where $\Delta$ = $\sqrt{(A/K_\mathrm{eff})}$ is the DW width. We used the values of $K_\mathrm{eff}$ for a given thickness of Pt from our previous work \cite{ogrodnik_study_2021}, while A = 16 pJ/m is the exchange stiffness \cite{emori_current-driven_2013}. The calculated $\Delta$ is in the range of 4 to 9 nm. This range agrees with data commonly reported in the literature \cite{emori_current-driven_2013, hrabec_measuring_2014}.

In order to verify the result for the $D_{\mathrm{eff}}$ constant, we recalculated $\Delta f$ based on experimentally derived values of $D_{\mathrm{eff}}$ using the SW model presented in Methods section. In our previous research\cite{ogrodnik_study_2021}, we showed that in the considered thicknesses of Pt, there are large variations in IEC as well as anisotropies in the top and bottom Co layers. Such a procedure provided insight into the reason for the dependence of $D_{\mathrm{eff}}(d_{\mathrm{Pt}})$ and its reliability in a trilayer system. As shown in Fig.\ref{fig:figure_dmi}(b), the calculated $\Delta f$ agrees well with the experimental dependence. The result was obtained for parametrized anisotropy constants and IEC as in Ref.\cite{ogrodnik_study_2021}.  Importantly, our calculations showed that the dependence holds even if anisotropies and IEC are fixed with a constant value, while DMI fields are the only parameters varying with Pt thickness. This means that the $\Delta f (t_{\mathrm{Pt}})$ dependence originates only from the change of DMI fields at interfaces, and not from the variations of other parameters. Therefore, the experimentally derived $D_{\mathrm{eff}}$ is reliable. Moreover, the calculations indicate that the DMI fields at both interfaces exhibit either similar magnitudes or, in the presence of asymmetry ($\chi=0$), the asymmetry is of a nature such that the average of these two DMI fields closely approximates $D_{\mathrm{eff}}$.

In Fig.\ref{fig:figure_dmi}(b), the value of $D_\mathrm{eff}$ is inversely correlated with the IEC, which, in turn, varies with the thickness of the Pt spacer (the dependence of IEC on Pt thickness is shown in Fig.\ref{fig:figure_7}(f)). This correlation suggests that changes in the IEC may be associated with adjustments in the DMI. Since the DMI arises from the exchange interaction between adjacent ferromagnetic spins, it implies that the IEC may influence this type of interaction. As the Pt thickness increases, the IEC interaction tends to weaken, potentially allowing the DMI coupling to play a more prominent role.
%is inversely related to the IEC coupling, which in turn depends on the thickness of the Pt spacer, showing that the DMI can be adjusted by interlayer exchange coupling. Given that the DMI comes from the exchange interaction between two adjacent FM spins, this suggests that the IEC exchange coupling results in an attenuation of this type of interaction. As the Pt thickness increases, the IEC interaction becomes weaker and the DMI coupling begins to play an increasingly important role.

\begin{figure}[h!]
\centering

\includegraphics[width=0.95\textwidth]{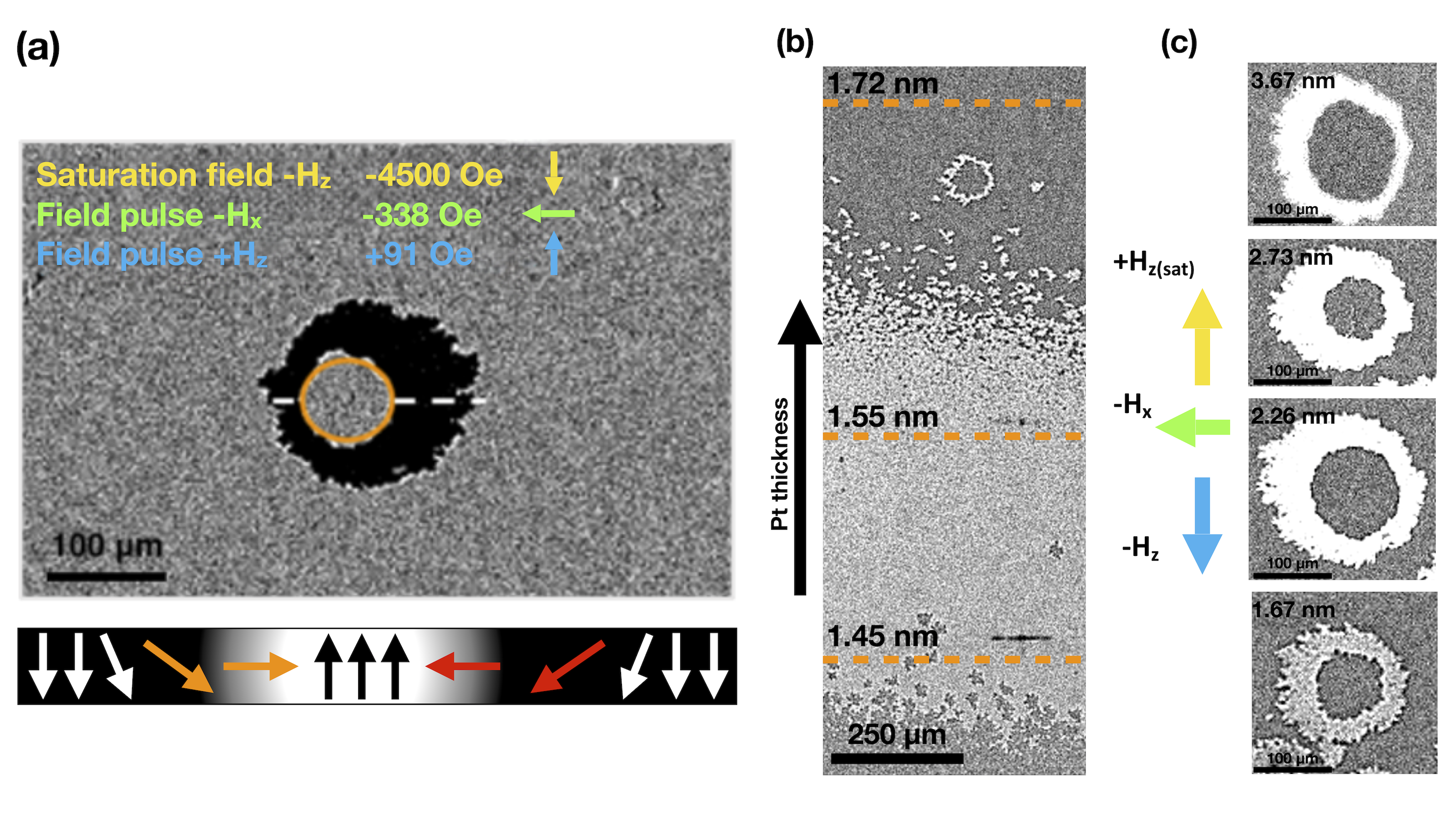}

\caption{(a) p-MOKE difference image showing the growth of bubble domains under the influence of the in-plane magnetic field ($H_{\mathrm{x}}$) and the out-of-plane magnetic field ($H_{\mathrm{z}}$) in the Co(1)/Pt(2.2)/Co(1) system. The applied field values and directions are illustrated in the accompanying image. The initial position of the bubble domain is indicated by an orange ring. Schematic representation of the lateral magnetization profile along the dashed white line is depicted in the image. The orange and red arrows denote the anticlockwise chirality of the N-DWs (N\'eel domain walls). (b) p-MOKE difference image of the continuous wedge layer of the studied system under the influence of $H_{\mathrm{x}}$ and $H_{\mathrm{z}}$ magnetic fields, where the orange dashed lines indicate the constant Pt thickness within the wedge. (c) Example p-MOKE differential images of the p-MOKE bubble domain structure for several Pt thicknesses located in regions II-IV shows the change in roughness of the bubble domain wall. Magnetic fields are applied in opposite directions to (a). }
\label{fig:chirality}
\end{figure}
We also investigated how in-plane ($H_{\mathrm{x}}$) and out-of-plane ($H_{\mathrm{z}}$) magnetic field pulses affect the expansion of bubble-like domains in a Co/Pt/Co films induced at $t_{\mathrm{Pt}}$ = 2.2 nm (Fig.\ref{fig:chirality}(a)). We have shown, like others \cite{kowacz_strong_2022} that the DMI causes asymmetric DW motion, as the external $H_{\mathrm{x}}$ field modifies the energy of Néel-type DWs (N-DWs) differently depending on their core magnetization direction. 
By applying a $H_{\mathrm{z}}$ saturation field, then tuning the  $H_{\mathrm{x}}$ and $H_{\mathrm{z}}$ fields and generating successive field pulses, we controlled the asymmetry of domain growth and determined the chirality of Co/Pt/Co. Our results show that the N-DWs have anticlockwise (ACW) spin configuration, indicating a positive DMI constant ($D_{\mathrm{eff}}$ > 0). 
A differential p-MOKE image of a continuous Co(1)/Pt(1.38-1.72)/Co(1) sample, subjected to pulsed $H_\mathrm{x}$ and $H_\mathrm{z}$, provides compelling evidence of the substantial impact of IEC magnitude on the size of the observed domain structure (see Fig.\ref{fig:chirality}(b)). The investigated Pt thicknesses fall within the range where the most significant reduction in coupling with increasing Pt thickness is observed.

In the range of the Pt thickness from 1.38 to 1.55 nm (Fig.\ref{fig:chirality}(b)), a remarkably fine-grained domain structure is observed, which progressively enlarges as the IEC decreases. At approximately 1.62 nm thickness, the structure transitions to a single bubble form. Furthermore, the influence of IEC on the shape of the domain wall within the single bubble domain structure was also observed (see Fig.\ref{fig:chirality}(c)). For high coupling ($t_{\mathrm{Pt}}$ = 1.67 nm), the domain wall exhibits a highly jagged character, while it becomes progressively smoother as the IEC strength decreases.

\subsection*{Current-induced magnetization switching \label{sec:SWITCH}}
\begin{figure}[ht!]
\centering
\includegraphics[width=1\textwidth]{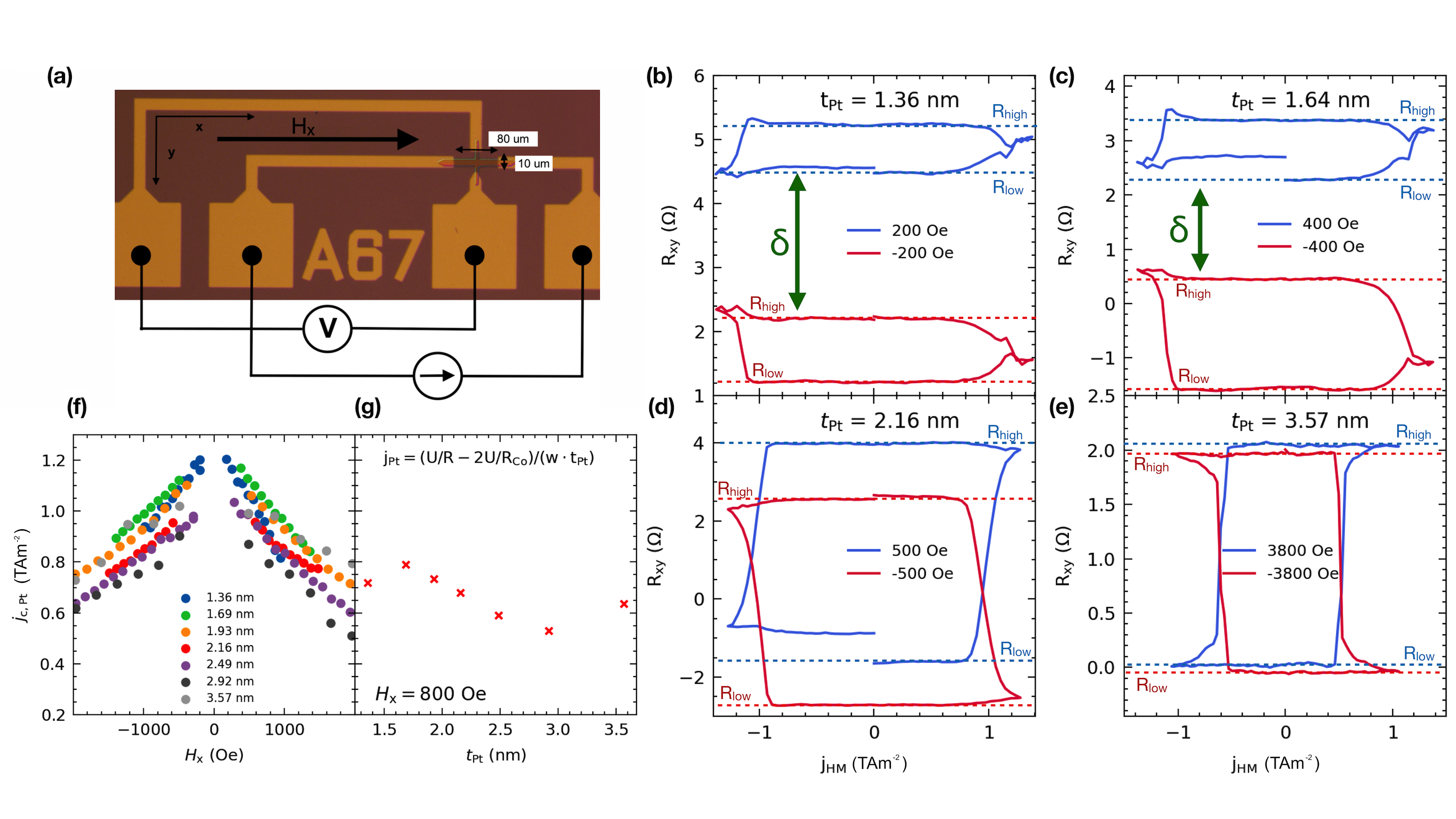}
\caption{(a) Device used for CIMS measurements. CIMS loops for devices from regions: (b)-(c) II, (d) III, and (e) IV. Blue and red dashed lines indicate the resistance levels of the bottom and top Co layers, respectively. The blue and red solid lines indicate CIMS loops for positive and negative magnetic fields, respectively. In accordance with the macrospin model, we denoted the gap between loops by $\delta$. Critical switching current density as a function of the external magnetic field ($H_\mathrm{x}$) for samples from regions II and III (f), critical current density ($j_{\mathrm{c,Pt}}$) as a function of Pt thickness (g). }
\label{fig:figure1}
\end{figure}

After determining the IEC and DMI dependence on Pt thickness, we now turn to the switching mechnism in the Co/Pt/Co system. We measured the anomalous Hall effect (AHE) to observe magnetization switching between two stable high and low resistance states of the current-pulsed loop. CIMS takes place in regions II-IV only, where at least one layer is perpendicularly magnetized in the remanent state. To achieve magnetization saturation, the samples were subjected to a large magnetic field in the $-z$ direction. Then, to drive magnetization switching, we applied a sequence of $1$ ms voltage pulses, with a pulse spacing of $2$ ms in the $x$ direction. The pulse amplitude was swept from $0$ V to a maximum positive value ($+V_{\mathrm{max}}$), then to the maximum negative value ($-V_{\mathrm{max}}$), and then back to $0$ V. Simultaneously, we measured the transverse voltage ($V_{\mathrm{xy}}$) in the presence of an in-plane magnetic field $H_{\mathrm{x}}$, which is co-linear to the current direction. The measurement setup is presented schematically in Fig.\ref{fig:figure1}(a).

The in-plane magnetic field $H_{\mathrm{x}}$ was changed sequentially after each CIMS loop in the wide range of $\pm 7~ \si{kOe}$. As a result, we obtained a set of CIMS loops in different $H_{\mathrm{x}}$ for representative Pt thicknesses from regions II to IV, and examples are plotted in Fig.\ref{fig:figure1}(b-e). 

As shown in Figs.\ref{fig:figure1}(b) and (c), experimentally obtained CIMS loops measured at positive and negative magnetic fields are clearly separated in regions II and III. Both stable resistance states of the CIMS loops have a higher resistance for $+H_{\mathrm{x}}$ (blue loop) than those measured for $-H_{\mathrm{x}}$ (red loop). When the direction of the magnetic field changes from +x to -x, we observed a smooth transition from the high-resistance loop to the low-resistance loop. For the thicker Pt spacer in Fig.\ref{fig:figure1}(d) ($t_{\mathrm{Pt}} = 1.64~\si{nm}$) separation gap becomes smaller compared to the sample of Pt = 1.36 nm thick (Fig.\ref{fig:figure1}(b). In region III ($t_{\mathrm{Pt}}= 2.16$ nm) (Fig.\ref{fig:figure1}(d), however, the four resistance states can still be observed. In the case of the thickest Pt ($t_\mathrm{Pt} = 3.57$ nm), for which only one Co layer exhibits perpendicular anisotropy, the separation gap disappears. Regardless of the direction of $H_{\mathrm{x}}$ only two resistance states exist, as in the case of the HM/FM bilayer (not shown here). \cite{miron_perpendicular_2011, liu_current-induced_2012,yu_switching_2014, grochot_current-induced_2021}

Subsequently, we performed an analysis of the critical current densities ($j_\mathrm{c,Pt}$) required to switch magnetization. For this purpose, the dependence of $j_\mathrm{c,Pt}$ through Pt was plotted as a function of the applied external magnetic field ($H_{\mathrm{x}}$). As demonstrated in Fig.\ref{fig:figure1}(f) for $H_{\mathrm{x}}$ $\ll$ $H_{\mathrm{k,eff}}$, the experimental dependencies measured in all devices are linear, which remains consistent with Ref.\citen{Lee2013}.

In Fig.\ref{fig:figure1}(g) we show the $j_{\mathrm{c,Pt}}$ dependence on the Pt layer thickness. The critical $j_{\mathrm{c,Pt}}$ decreases linearly in a wide range of Pt thickness, from 1.6 to 3 nm, when it reaches its lowest value. However, for the thinnest and thickest Pt layer, $j_{\mathrm{c,Pt}}$ deviates from the linear dependence by slightly dropping and rising, respectively. The highest values of the critical current amplitude required for switching are found for elements with a small thickness of Pt and then decrease linearly slightly to a value of approximately $0.5~\mathrm{TAm^{-2}}$ for the element with $t_{\mathrm{Pt}} = 2.92$ nm. 

\begin{figure}[ht!]
\centering
\includegraphics[width=0.7\linewidth]{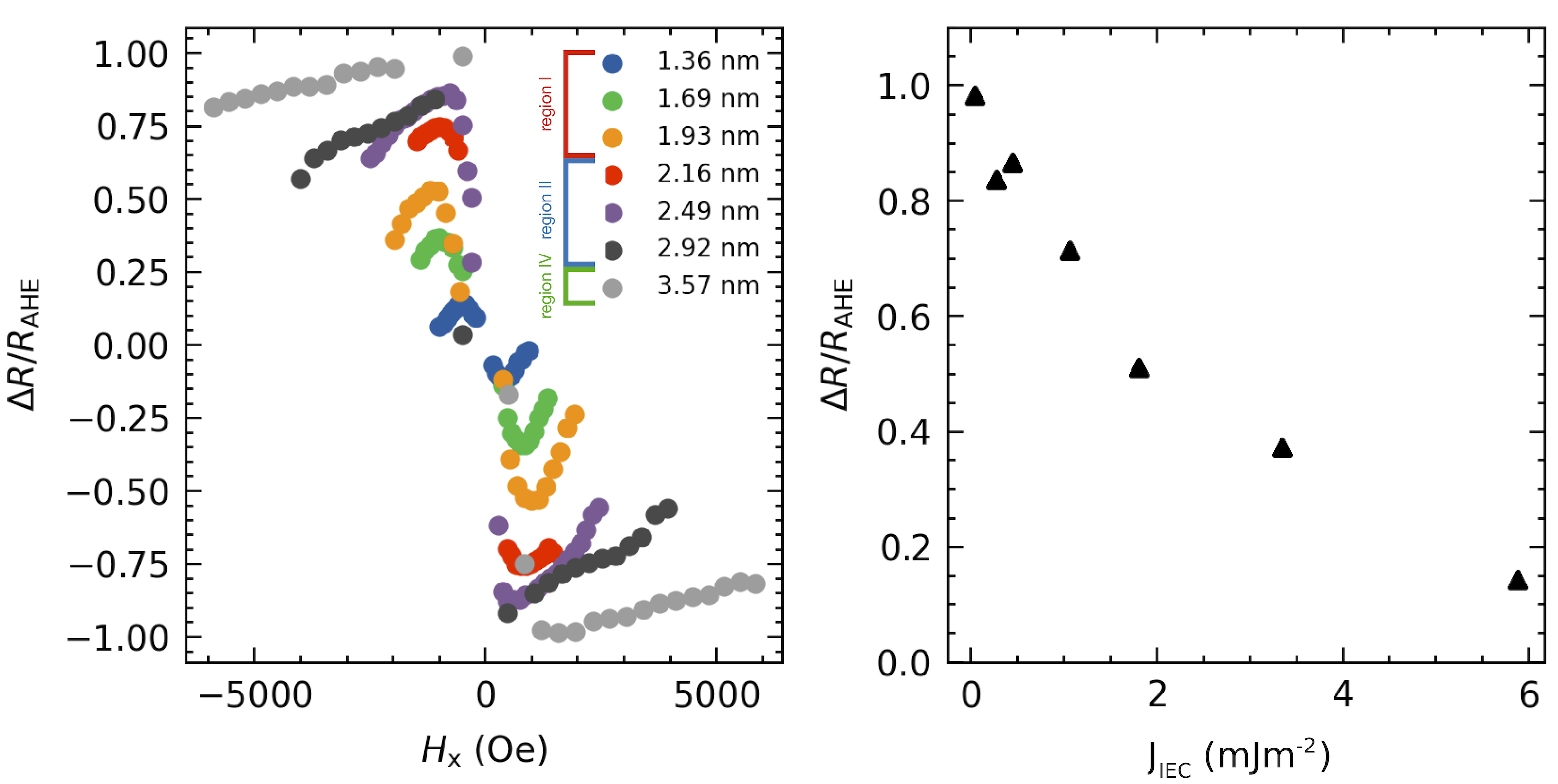}
\caption{(a) $\Delta R$/$R_{\mathrm{AHE}}$ ratio as a function of the external magnetic field for elements with different Pt layer thicknesses. (b) The maximum values of $\Delta R$/$R_{\mathrm{AHE}}$ ratio as a function of IEC.}
\label{fig:dr}
\end{figure}

One of the most important components of the current-induced magnetization switching process, from an application point of view, is the difference between high and low resistance levels of the current switching loop, denoted $\Delta R = R_{\mathrm{high}}-R_{\mathrm{low}}$. 
In the case of the devices studied, the amplitude $\Delta R$ depends on the thickness of the Pt spacer and, therefore, on the magnitude of the IEC. Elements with a thin Pt layer (region II and III) and, thus, with strong coupling, exhibit small values of $\Delta R$, reaching a $\Delta R$/$\Delta R_{\mathrm{AHE}}$ value of $0.8$ (Fig.\ref{fig:dr}(a). 
\begin{figure}[h!]
\centering
\includegraphics[width=8cm]{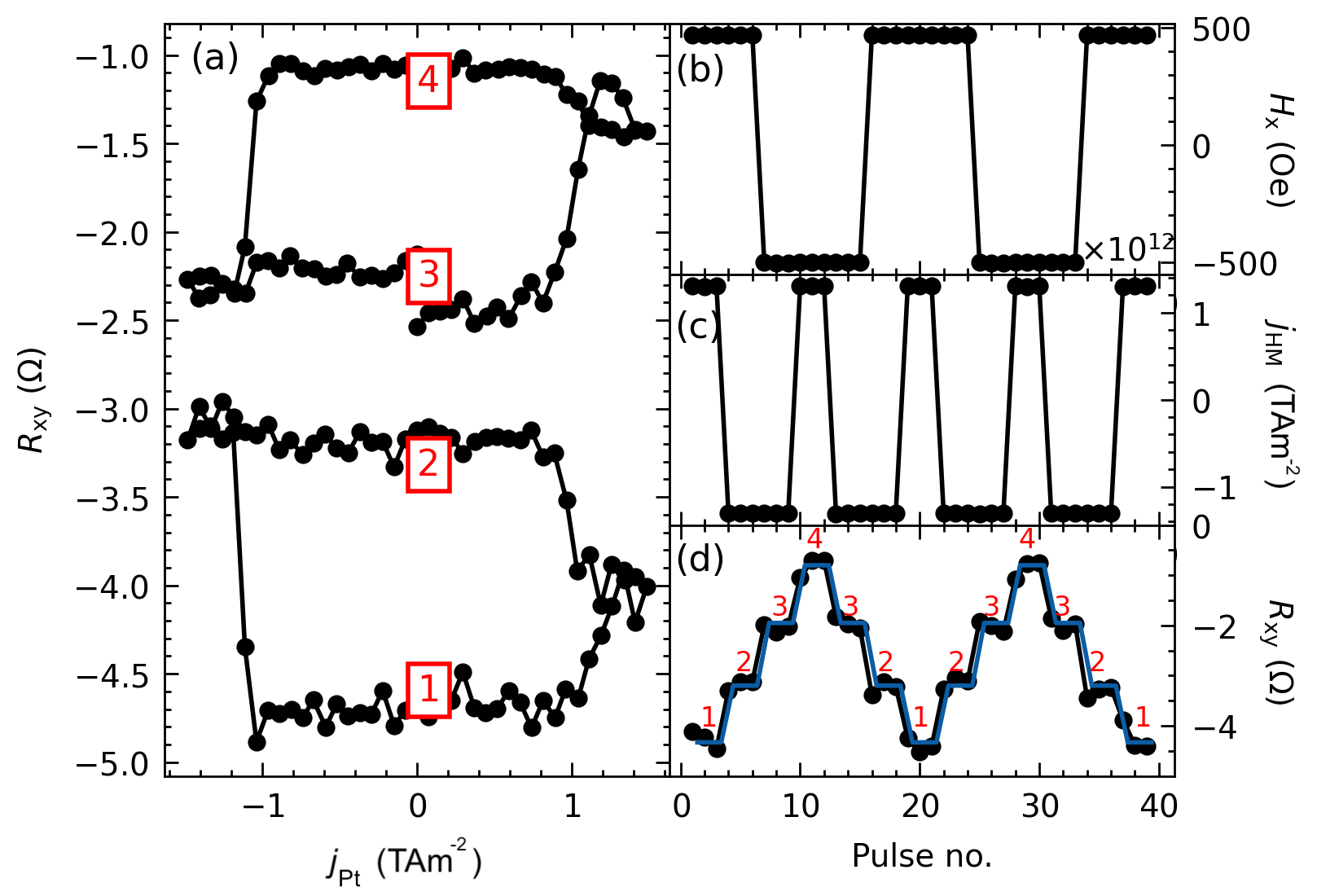}
\caption{Four stable resistance states for the sample from region II ($t_{\mathrm{Pt}}$ = $1.55$ nm (a), obtained by manipulating the magnitude of the current pulse (b) and the external magnetic field (c). Resistance levels in (a) correspond to the resistance levels of CIMS loops in (d).}
\label{fig:multilevel_switching}
\end{figure}
The fact that the amplitude ratio does not reach the maximum value ($\Delta R$/$\Delta$ $R_{\mathrm{AHE}}$ < 1) indicates a domain-specific origin of switching and therefore the magnetic domains persist in remanence. Elements in region IV, where the coupling is negligible (Fig.\ref{fig:dr}(b), show $\Delta R$/$\Delta R_{\mathrm{AHE}}$ values close to $1$, suggesting that the switching is practically single-domain and only Co layer with perpendicular anisotropy switches.

Therefore, to investigate multilevel switching, we focused on $t_{\mathrm{Pt}}$ = $1.55$ nm (region II). For this thickness, we chose two values of the external magnetic field (+0.5 kOe and -0.5 kOe) at which both loops exhibit a significant amplitude ($\Delta R$) and are completely separated. %For $t_{\mathrm{Pt}}$ = $1.55$ nm (region II), we chose two values ($+0.5$ $\si{kOe}$ and $-0.5$ $\si{kOe}$) of the external magnetic field for which both loops show significant amplitude ($\Delta R$) and are completely separated. 
As shown in Fig.\ref{fig:multilevel_switching}(a), there are four different resistance states. On the basis of the switching loops, we determined the critical current densities ($j_{\mathrm{c,Pt}}$) of +$1.29$ $\mathrm{TAm^{-2}}$  and -$1.29$ $\mathrm{TAm^{-2}}$ needed to switch the magnetization at $\pm 0.5$ kOe. Then, to switch the resistance between four well-separated levels, we applied both current pulses of $\pm I_\mathrm{c}$ amplitude and the magnetic field of magnitude of $\pm 0.5$ kOe (Fig. \ref{fig:multilevel_switching}(a). By carefully choosing the combination of signs $H_\mathrm{x}$, shown in the Fig.\ref{fig:multilevel_switching}(b), and $j_\mathrm{c}$, given in Fig.\ref{fig:multilevel_switching}(c), we obtained a ladder-shaped waveform of resistance of the system (Fig.\ref{fig:multilevel_switching}(d). The procedure of tuning the switching pulse duration and its amplitude for an arbitrary field allows the system to be set in a single well-defined resistance state and therefore to store considerably more information in a single memory cell.

\subsection*{Domain mechanism of multilevel switching}
\begin{figure}[h!]
\centering

\includegraphics[width=1\textwidth]{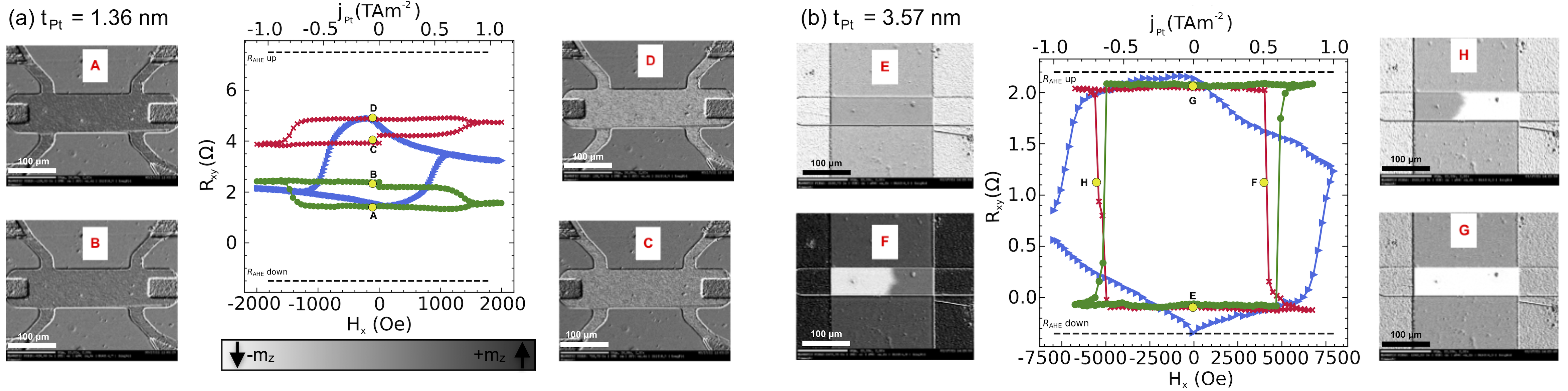}

\caption{CIMS loops for (a) $t_{\mathrm{Pt}} = 1.36$ nm thick element (region II) and (b) $t_{\mathrm{Pt}}
= 3.57$ nm thick element (region IV). The CIMS loops (red and green) were obtained in fields $H_\mathrm{x}$ of $+0.8$ $\si{kOe}$ and $-0.8$ $\si{kOe}$ in (a) and $+3.8$ $\si{kOe}$ and $-3.8$ $\si{kOe}$ in (b), respectively. The blue triangles indicates the $R_{\mathrm{xy}}$ loops measured in the $H_\mathrm{x}$ field. The letters (A-H) in figures (a) and (b) indicate the relevant p-MOKE images labeled with the same letter. }
\label{fig:moke_pictures}
\end{figure}

We qualitatively explain the observed CIMS loops in terms of magnetic domains in microstrips of Hall-bar devices. For this purpose, selected Hall-bars from each region of Pt thickness were imaged with p-MOKE while the $H_\mathrm{x}$ field was applied. The resistance $R_{\mathrm{xy}}$ vs. $H_{\mathrm{x}}$ represents the reversal magnetization process by blue loops (Fig.\ref{fig:moke_pictures} a,b). It enabled us to relate the change of magnetic domain structure with the resistance level measured during CIMS. Firstly, the magnetization of each Hall bar was saturated with $H_\mathrm{z}$ field to the lowest resistance state, indicated by A and E in Figs.\ref{fig:moke_pictures}(a) and (b), respectively. Therefore, it was possible to assign images of the domain structure to the corresponding resistances in the current switching loops, as shown in Figs.\ref{fig:moke_pictures}(a) and (b). 

We also repeated this procedure for the perpendicular field $H_\mathrm{z}$ and in this case the magnetization reversal was performed by a single-domain wall motion, represented by a rectangular AHE loop (not shown).

The p-MOKE images reveal that the resistances within the CIMS loops correspond to a very fine-grained domain structure (Fig.\ref{fig:moke_pictures}(a) A-D) in the Hall bar with thickness $t_{\mathrm{Pt}} = 1.36$ nm (region II) in the remanence state $H$=0 ($j$=0). When current-induced SOT switches the magnetization, a number of domains change their state to the opposite. The new distribution of magnetic domains results in an intermediate state (yellow dots B and C) placed between the two extremes marked in Fig.\ref{fig:moke_pictures}(a) with letters A and D. This transition of magnetic domains can be observed as a change in the gray color level of elements marked A and B (or C and D) in Fig.\ref{fig:moke_pictures}(a). The smooth shape of the CIMS loops in this region confirms the fine-grained magnetic domain switching mechanism.

The opposite behavior occurs in the region IV element with a Pt thickness of $3.57$ nm, where we observed a complete magnetization reversal driven by a current-induced domain-wall motion. This behavior demonstrates itself as a perfectly rectangular
shape of the CIMS loops with only two stable resistance states for both directions of the magnetic field (+$H_\mathrm{x}$ and -$H_\mathrm{x}$) (Fig.\ref{fig:moke_pictures}(b)

Matching the CIMS resistances involved generating the fine-grained structure visible in Fig.\ref{fig:moke_pictures}(a). It was achieved by, first, saturating the sample with a perpendicular field ($H_\mathrm{z}$), then applying a field $H_\mathrm{x}$ of about $10$ $\si{kOe}$, and then gradually reducing its value to approximately $1$ $\si{kOe}$. As a result, the field-free resistance of the system is not equal to the high-resistance state of the AHE loop due to the uneven distribution of the $m_\mathrm{z}$ components of the magnetic domains in both Co layers. This condition is presented in Fig.\ref{fig:moke_pictures}(a), where there is a predominance of domains with a +$m_\mathrm{z}$ component at remanence. Reapplying a small $H_\mathrm{x}$ field generates a $m_\mathrm{x}$ component parallel to the direction of the magnetic field in both Co layers, the existence of which is necessary for the switching of magnetization by the spin-polarized current. 

\begin{figure}[h!]
\centering
\includegraphics[width=0.8\textwidth]{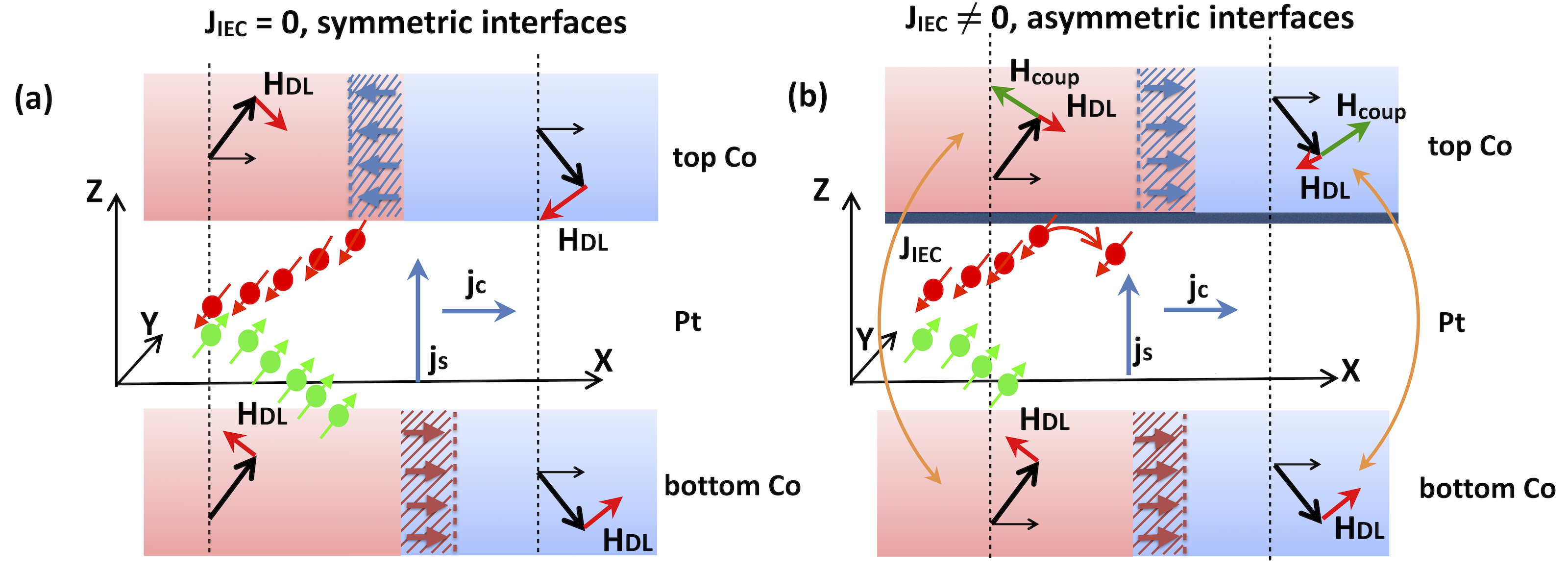}
\caption{Mechanism of SOT-CIMS in two cases: With no coupling ($J_\mathrm{IEC}$=$0$) and symmetric Co/Pt – Pt/Co interfaces (a) and in the presence of strong coupling and asymmetric interfaces (top interface with less spin-transparency is marked as solid navy blue layer) (b). The red(blue) areas represent magnetic domains with average +$m_z$(-$m_z$) components. The dashed areas together with horizontal arrows indicate the change in domain size under the $H_\mathrm{DL}$ SOT components (thick red arrows) and ferromagnetic coupling $H_\mathrm{coup}$ field (thick green arrows, thin solid orange arrows show the coupling between Co layers). The spin current with polarization +$\mathbf{e}_y$ (-$\mathbf{e}_y$) is depicted as red (green) bold points with arrows.   
}
\label{fig:schemat_przelaczania}
\end{figure}

The following scenario explains the behavior of CIMS: in the top Co layer, the current-induced SOT damping-like effective field ($\mathbf{H}_\textrm{DL} \sim -\mathbf{m} \times \mathbf{e}_y$) acts oppositely on domains with $+m_z$ and $-m_z$ components, i.e., for positive currents, +$H_\textrm{DL}$ forces $+m_z$ domains to switch, while $-H_\textrm{DL}$ pushes $-m_z$ domains back to the perpendicular direction. 
On the other hand, the spin current flowing into the bottom Co layer has the opposite sign. Therefore, SOT stabilizes the $+m_z$ domains while switching the $-m_z$ domains in this layer (Fig.\ref{fig:schemat_przelaczania}(a). Firstly we discuss the effect of the SOT field on an uncoupled and fully symmetric trilayer as depicted in Fig.\ref{fig:schemat_przelaczania}(a). 
When the current pulse reaches a critical amplitude value, each of the Co layers can switch only partially. However, we note that, in a fully symmetric and uncoupled case, the SOT would not result in a resistance change. Then, the increase of $-m_z$ in one layer would be balanced by the increase of $+m_z$ in the second layer, which is illustrated in Fig.\ref{fig:schemat_przelaczania}(a) with horizontal arrows pointing in opposite directions. However, thin Pt devices (region II) are far from the symmetric case (Ref.\citen{ogrodnik_study_2021}). In \ref{fig:schemat_przelaczania}(b) we show the scenerio of the coupled and asymmetric case.  The top and bottom interfaces differ, and therefore the magnitudes of effective $H_\textrm{DL}$ fields acting on each Co are not equal, as shown in our previous paper \cite{ogrodnik_study_2021} . Moreover, the magnetic anisotropies in both layers are different, and a large ferromagnetic IEC is present in this region.\cite{ogrodnik_study_2021} For this reason, when the lower Co layer switches, the ferromagnetic coupling forces the magnetization of the upper Co layer to switch as well. 

The switching process results in a higher transversal resistance related to a larger number of domains with magnetization pointing in the +z, rather than the -z direction, in both Co layers.

The described mechanism is consistent with the dependence of the critical current density ($j_\mathrm{c,Pt}$) on the thickness of Pt presented in Fig.\ref{fig:figure1}(g).
$j_\mathrm{c,Pt}$ decreases for the Pt thickness, ranging from $1.7$ to $3.0$ nm (regions II and III). This decrease is due to a more efficient SOT for the thicker Pt layer.\cite{Nguyen_2016} However, for the thinnest Pt in region II ($1.3$ nm), $j_\mathrm{c,Pt}$ drops by about $\Delta j_\mathrm{c}$ = $0.10$ $\mathrm{TAm^{-2}}$ ). Similarly, for the thickest Pt in region IV ($3.57$ nm), the critical current abruptly increases approximately ($\Delta j_\mathrm{c}$ =  $0.17$ $\mathrm{TAm^{-2}}$). 
The deviations from the linear dependence are correlated with very strong coupling (for the thinnest Pt) and negligible coupling (for the thickest Pt). The switching in the thin Pt case relies on the magnetization reversal in both Co layers. These two layers have different anisotropy fields ($H_{\mathrm{k,eff}}$ (top) > 0, $H_{\mathrm{k,eff}}$ (bottom)<0). \cite{ogrodnik_study_2021} It means that the bottom layer is more susceptible to torque from the SOT effect. Therefore, when the IEC field is strong enough, it easily overcomes $H_{\mathrm{k,eff}}$ in the top layer, allowing it to switch at a lower current (SOT). Then, both Co layers are magnetically stiff and behave somewhat like one layer with the effective anisotropy: $H_{\mathrm{k,eff}}$(top)>$H_{\mathrm{k,eff}}$> $H_{\mathrm{k,eff}}$(bottom). 

For the intermediate IEC (border of regions II and III), the bottom layer is still more switchable, but the coupling does not provide the top layer with enough torque to switch. Both layers become less magnetically stiff, so more current (more SOT) is needed to switch both of them. 

The bottom layer magnetization is in-plane when the coupling becomes negligible (region IV). It means that the SOT only switches the top layer with higher anisotropy ($H_{\mathrm{k,eff}}$(top)). Therefore, the critical current rises despite the thick Pt and large SOT.

\subsection*{Macrospin and 1D domain wall simulation for multilevel switching}
\label{sec:model}

We attempted to reproduce the experimental results with the simplest possible model. To this end, we employed two macrospin-based models: the LLGS model and the 1D domain wall model (see Methods for details). Reproduction of the hysteresis gap under field reversal is achieved by a small modification of the resistance model of Kim et al. \cite{kim2016} For simplicity, we neglected $\Delta R_\mathrm{SMR_{xy}}$ and $\Delta R_\mathrm{AMR_{xy}}$ due to their small contributions, since the term $(\Delta R_\mathrm{SMR_{xy}} + \Delta R_\mathrm{AMR_{xy}})m_x m_y$ tends to vanish in macrospin simulations as the $m_y$ component becomes negligible. Based on the previous work\cite{ogrodnik_study_2021}, we posited that the contributions of the top and bottom FM layers are not equal due to the different intermixing of Co and Pt atoms at both interfaces. This assumption is accounted for in the model as an additional resistance asymmetry parameter $\beta$, leading to $R_\mathrm{xy}$ being computed as:
\begin{equation}
    \label{eq:Rxy}
    R_\mathrm{xy} \approx R_\mathrm{xy0}^{(1)} +  R_\mathrm{xy0}^{(2)} + \frac{1}{2}\kappa\Delta R_\mathrm{AHE}(m_z^{(1)} + \beta m_z^{(2)})
\end{equation}
where the superscript refers to the top $(1)$ or the bottom $(2)$ FM layer. The parameter $\beta$ ranges from 0 exclusive (asymmetric interfaces) to 1 inclusive for entirely symmetric interfaces. The dimensionless parameter $\kappa$ effectively corrects the amplitude $R_\mathrm{xy}$ for the multidomain behavior in regions II through III, which is necessary due to limitations of the macrospin model.

\begin{figure}[ht!]
    \centering
    \includegraphics[width=8.3cm]{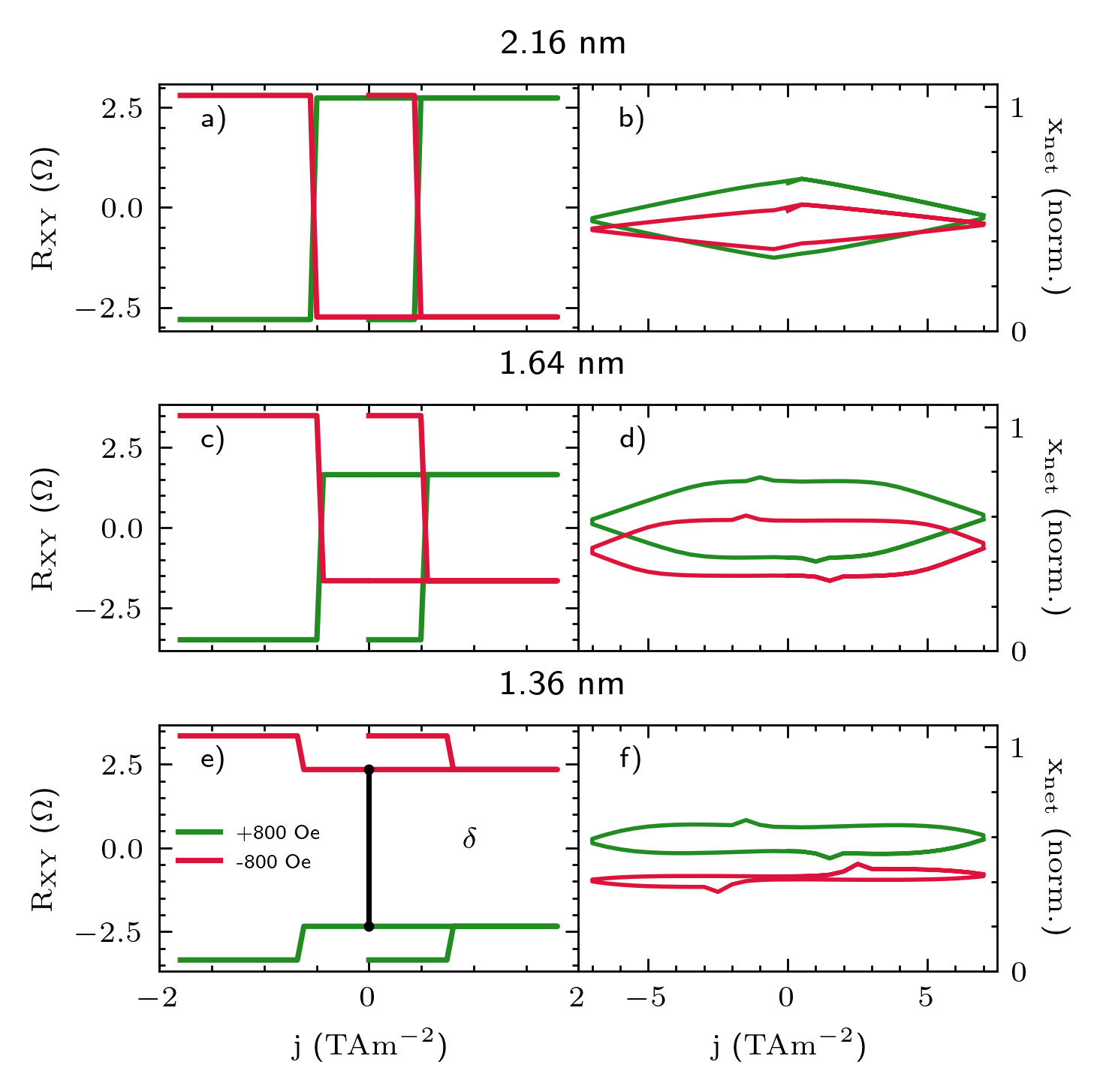}
    \caption{(a, c, e) CIMS multilevel switching for a range of thicknesses. The external field was the same for all simulations with $H_\textrm{x} = \pm 0.8~ \si{~kOe}$. An example of $\delta$ separation gap was marked in (e).
    (b, d, f) Multilevel switching for a range of thicknesses in the 1D domain wall model.}
    \label{fig:switching-macro}
\end{figure}

In the simplified domain wall motion model, we rephrase the resistance model as the net distance traveled by the domains in the two layers, $x_\mathrm{net}$. Assuming a finite strip length, the domain can eventually reach either the left or right edge of the strip. If the domain separates the up ($m_z > 0$) and down ($m_z < 0$) states, then the more the domain moved to the right, respective to the center marked at $x = 0$, the higher the resistance state $R_\mathrm{xy}$ would be, as per Eq.\ref{eq:Rxy}. 
Figure \ref{fig:switching-macro} presents the simulated results of the macrospin depicted in the left side panels (a, c, e) and the 1D domain wall motion model shown in the right side panels (b, d, f). The absence of a proper hysteresis shape in the 1D domain wall simulation is attributed to the lack of an edge field, which typically slows down the domain as it approaches the sample's edge. Instead, we simply simulated reaching the edge by the current reversal; i.e. in our model the domain always reaches the edge of the sample for the minimum and the maximum current. In Fig.\ref{fig:switching-macro} two main features of the experimental findings are preserved; first, the gap separation is achieved and it shrinks as the Pt layer becomes thicker (and the coupling decreases). This shrinkage corresponds to an increase in the value of $\beta$, indicating a smaller interfacial asymmetry. Second, the critical currents decrease with the growing thickness of the HM layer, which our model replicates with an adequate increase in the field- and damping-like torque values.

Importantly, the simulations demonstrated that DMI does not alter the qualitative outcome. Although DMI influences the amplitude of the $H_{\mathrm{x}}$ field, depending on its orientation, it does not sufficiently, by itself, explain the distinct separation of the hysteresis states during the field reversal. However, its presence may slightly reduce the gap between the two loops. We conclude that the asymmetry of the two Co/Pt and Pt/Co interfaces is crucial for obtaining the multilevel switching behavior.  

In summary, the experimentally observed multilevel switching primarily originates from the differences in spin-transparency at the interfaces. This feature enables effective IEC-mediated domain structure switching in both layers, as demonstrated by MOKE imaging, electrical CIMS measurements and related simulations. Additionally, it is notable that the mechanism of IEC-mediated switching aligns with the critical current dependence on Pt thickness. While the IEC can potentially tailor DMI if it holds sufficient strength to alter the magnetization distributions at interfaces, its direct impact on the multilevel switching itself seems relatively limited. Therefore, both DMI and IEC as well as the asymmetry of the HM/FM interfaces are necessary to design a structure with more than two transversal resistance states.

\section*{Methods}

\subsection*{Device fabrication}
\begin{figure}[h!]
\centering
\includegraphics[width=1\linewidth]{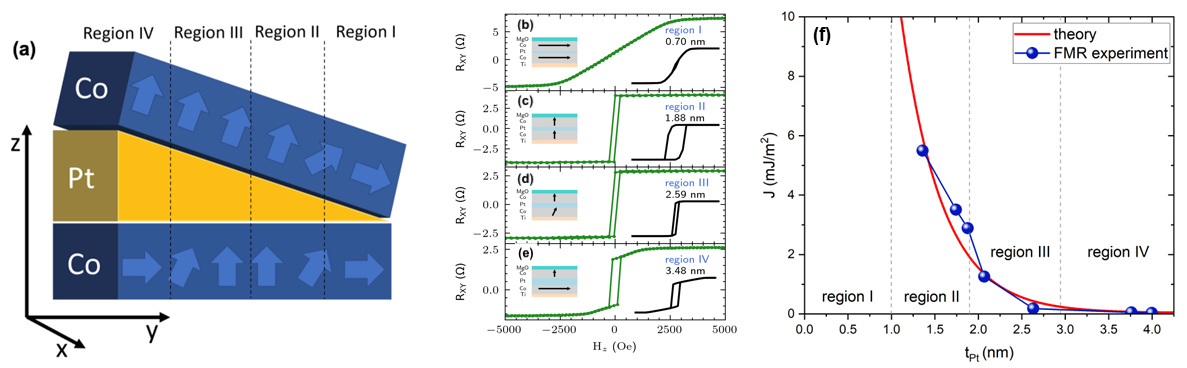}
\caption{(a) Cross-section through the studied heterostructure. The blue arrows depict the direction of magnetization vectors in both ferromagnetic Co layers in specified Pt thickness regions. The dashed lines indicate the border of each region. (b)-(e) AHE loops for Hall-bar devices with different thicknesses of the Pt spacing layer. The solid black lines in the inset denote the simulated AHE loops using the model described in Sect. Macrospin models. The depicted diagrams of multilayer cross sections for all regions indicate the direction of magnetizations of magnetic layers at
remanence. (f) Interlayer coupling derived from the macrospin simulations of the spin diode FMR spectra.}
\label{fig:figure_7}
\end{figure}

The continuous, wedge-shaped FM/HM/FM heterostructure was deposited using a magnetron sputtering technique on the $20~ \times ~ 20$ $\mathrm{mm^2}$ $\mathrm{Si/SiO_2}$ substrate at room temperature and under the same conditions as in Ref.\citen{ogrodnik_study_2021}. The sample cross-section scheme and the coordinate system used are shown in Fig.\ref{fig:figure_7}(a). Layers are ordered as follows: Si/$\mathrm{SiO_2}$/Ti(2)/Co(1)/Pt(0-4)/Co(1)/MgO(2)/Ti(2) (thicknesses listed in parentheses are in nanometers). Both the bottom and the top Ti layers function as the buffer and the protection layer, respectively. They do not contribute to the studied phenomena as a result of their partial oxidation and small spin-orbit coupling. \cite{sule_asymmetric_2008, avci_interplay_2014, poh_direct_2021} After the deposition process, the sample was characterized by X-ray diffraction. We detected the presence of a face-centered cubic fcc(111) texture at the Pt/Co and Co/Pt interfaces and confirmed the existence of an asymmetry between these two interfaces. Details of the structural analysis of the studied samples are described in Ref.\citen{ogrodnik_study_2021}.

We performed X-ray reflectivity measurements (XRR) to precisely calibrate the thickness of each layer as a function of the position on the sample wedge. In doing so, we were able to precisely determine the thickness of the layers located at a specific position on the wedge of Pt. The variation in thickness of the Pt layers in the device was less than 0.006 nm, so the Pt thickness is constant throughout the device. The sample was nanopatterned by optical laser lithography, ion etching, and lift-off to a matrix of different sizes of Hall bar devices, which were optimized for the measurement techniques used. We used Hall bars of size $80$ x $10$ $\si{\mu m^2}$ for current-induced magnetization switching (CIMS) measurements, while resistance and magnetoresistance measurements were performed on $140 \times 20$ $\si{\mu m^2}$ devices using the 4-probe method. 

\subsection*{Anomalous Hall effect and effective anisotropies}
We measured the AHE for all elements along the Pt wedge. As a result, we obtained a set of AHE resistance loops as a function of the external magnetic field applied along the $z$ direction ($H_{\mathrm{z}}$). By analyzing their shapes, we could distinguish four regions of Pt thickness (marked regions I-IV) in which the AHE loops exhibit a similar shape (Fig.\ref{fig:figure_7}(b)-(e)). As shown in our previous work \cite{ogrodnik_study_2021}, in region I, the magnetizations of both Co layers are in-plane ($K_{\mathrm{eff}}$<0) and, as a consequence, AHE depends linearly on the magnetic field $H_\mathrm{z}$. Therefore, it is not possible to distinguish resistance states with AHE during CIMS in this region. Regions II and III were characterized by two Co layers magnetized perpendicularly to the sample plane in the remanent state and strong IEC (Fig.\ref{fig:figure_7}(f)) (both Co layers switch simultaneously), and as a result, the AHE hysteresis loops become rectangular, as demonstrated in Fig.\ref{fig:figure_7}(b),(c). Moving from region III to IV, the interlayer exchange coupling (IEC) decreases substantially as the Pt spacing layer thickness. \cite{ogrodnik_study_2021} Consequently, the top Co remains magnetized perpendicularly, whereas the bottom layer tends to be magnetized in the plane again.

\subsection*{Macrospin model}
\label{sec:model}
The model is based on Landau-Lifshitz-Gilbert-Slonczewski (LLGS) of the following form \cite{ralph_spin_2008,slonczewski_current-driven_1996,nguyen_spinorbit_2021}:
\begin{equation}
	  \frac{\textrm{d}\mathbf{m}}{\textrm{dt}} = -\gamma_0 \mathbf{m} \times \mathbf{H}_{\mathrm{eff}} + 
	  \alpha_\textrm{G} \mathbf{m}\times \frac{\textrm{d}\mathbf{m}}{\textrm{dt}}
	  -\gamma_0 H_\textrm{FL}(\mathbf{m} \times \mathbf{e}_y)
	  -\gamma_0 H_\textrm{DL}(\mathbf{m}\times\mathbf{m}\times \mathbf{e}_y)
\label{eq:llg-sot}
\end{equation}    
where $\mathbf{m} = \frac{\mathbf{M}}{M_s}$ is the normalized magnetization vector, with $M_\textrm{s}$ as magnetization saturation, $\alpha_\textrm{G}$ is the dimensionless Gilbert damping coefficient, $\gamma_{0}$ is the gyromagnetic ratio, $H_\textrm{DL}$ and $H_\textrm{FL}$ are damping-like and field-like torque amplitudes respectively, and $\mathbf{e}_y$ is the spin polarization vector in y direction. The $\mathbf{H}_\textrm{eff}$ is the effective field vector that includes contributions from the external magnetic field, anisotropy, IEC, and demagnetization energy.
For the reproduction of the experimental results, we used the open source package \textsc{cmtj} \cite{cmtj_2022}, taking the simulation parameters from Ref.\citen{ogrodnik_study_2021}. Small in-plane components of anisotropy help break the symmetry under an external field $H_{\mathrm{x}}$. In macrospin simulations, we take the Gilbert damping of $\alpha_\textrm{G} = 0.05$.

\subsection*{1D domain wall model}
The equations of domain wall motion are given by\cite{emori_current-driven_2013,yang_novel_2017}:
\begin{gather}
    (1 + \alpha^2) \dot{X} = \Delta (\Gamma_A + \alpha \Gamma_B) \nonumber \\
    (1 + \alpha^2) \dot{\phi} = -\alpha \Gamma_A + \Gamma_B
\end{gather}
where $X$ and $\phi$ are coordinates: position and angle, $\Delta$ is the domain width, $\Gamma_A$ contains effective field terms and $\Gamma_B$ contains non-conservative SHE field:
\begin{equation*}
    \Gamma_A = \gamma_0\left(-\frac{1}{2}H_k\sin2\phi + \frac{\pi}{2}H_x\sin\phi + \frac{\pi}{2}H_\textrm{DMI}\sin\phi + \frac{2J_\textrm{IEC}}{\mu_0 M_s t_\mathrm{FM}}\sin(\phi - \phi') \right)
\end{equation*}
\begin{equation*}
    \Gamma_B = \gamma_0\frac{\pi}{2}H_\textrm{SHE}\cos\phi
\end{equation*}
where $\phi$ and $\phi'$ denote the angles of DWs in the coupled layers\cite{yang_novel_2017}. We solved the equations numerically using the Runge-Kutta 45 method. 
%As indicated by the equations of motion, the inclusion of the DMI term does not change the qualitative result: It has the effect of increasing or decreasing the amplitude of the $H_\mathrm{x}$ field, depending on its orientation, creating an asymmetry that is not sufficient to explain the separation of the hysteresis states under field reversal, nevertheless its presence may decrease the gap between two loops. 
The magnitudes of the Dzyaloshinskii-Moriya interaction (DMI) were taken from Fig.\ref{fig:figure_dmi}.

\subsection*{Spin wave model\label{sec:SW}}
To compare the results with theoretical predictions, we employed a simplified model for the Damon-Eschbach (DE) mode of spin waves. As a starting point, we have used a SW model by Kuepferling et al.\cite{RevModPhys.95.015003}. Next, we extended the model to Co/Pt/Co trilayer systems by incorporating the Co interlayer coupling energy density:
$$E_{\mathrm{IEC}} = -J_{\mathrm{IEC}} \mathbf{m}_{\mathrm{1}}(\mathbf{r}) \cdot \mathbf{m}_{\mathrm{2}}(\mathbf{r})$$
where $\mathbf{m}_{\mathrm{1,2}}(\mathbf{r})$ is a space- and time-dependent magnetization in both layers, and $J_{\mathrm{IEC}}$ denotes IEC coupling amplitude.
Such an approach does not explicitly account for the coupling due to the dynamic dipolar field produced by the SW modes. It could be done as long as the interlayer coupling through Pt is much larger than that for the dipolar fields acting on each other. Despite the simplifications we made, the dipolar fields influence the dynamics of each Co layer separately, similarly as in Ref.\citen{RevModPhys.95.015003}. Our extension required considering two Co layers. Therefore, we assumed that the magnetizations in the two layers are in the form $\mathbf{m}_{(1,2)} = \mathbf{M}_{(1,2)}(\mathbf{r},t) = M_s (\delta m_x, 1, \delta m_z)$, where $\delta m_{(x,z)}$ are small deviations of the magnetizations saturated in the $y$ direction due to the SW DE mode. Therefore, the dynamical components $x$ and $z$ of magnetization had the following form: $\delta m_{(x,z)}(\mathbf{r},t) = \delta m_{(x,z)} e^{i (\mathbf{k} \cdot \mathbf{r} + \omega t)}$. 
Next, we linearized the Landau-Lifschitz(LL) equation:
$\frac{\textrm{d}\mathbf{m}_{(1,2)}}{\textrm{dt}} = -\gamma_0 \mathbf{m}_{(1,2)} \times \mathbf{H}_{\mathrm{eff,(1,2)}}$, 
where, similarly as in Eq.~\ref{eq:llg-sot}, $\mathbf{H}_{\mathrm{eff,(1,2)}}$ includes the external and anisotropy field as well as the micromagnetic spin-wave-induced fields related to demagnetization:
$\mathbf{H}_{\mathrm{d}} = -\textrm{Ms}\left(\delta \textrm{m}_{x (1,2)} \textrm{P}(|k|),0, \delta \textrm{m}_{z  (1,2)} (1 - \textrm{P}(|kd|) \right)$, where $\textrm{P}(|kd|) = 1 - \frac{1 - \textrm{e}^{-|kd|}}{|kd|}$ where $d$ represents the thickness of the Co layers, and $k$ denotes the wave vector length. Inter- (IEC) and intralayer exchange interactions are described by: 
$\mathbf{H}_{\textrm{IEC}} = \frac{J_\textrm{IEC}  }{\mu_0 M_s d}\mathbf{m}_{(2,1)}$ and 
$\mathbf{H}_{\textrm{ex}} = \frac{2A}{\mu_0 M_s} \nabla^2 \mathbf{m}_{(1,2)}$ with $A$ denoting the exchange constant, whereas the DMI fields $\mathbf{H}_{\mathrm{DMI,(1,2)}}$ are given by $\mathbf{H}_{\mathrm{DMI,(1,2)}} = \frac{2D_{1,2}}{\mu_0 M_s} \left(\frac{\partial m_{z,(1,2)}}{\partial x}, 0, -\frac{\partial m_{x,(1,2)}}{\partial x}\right)$ characterized by two DMI constants corresponding to each Co/Pt interface. Moreover, we parametrized DMI constants in the following way: $\mathrm{D_1} = \mathrm{D_{eff}} - \chi$ and $\mathrm{D_2} = \mathrm{D_{eff}} + \chi$, maintaining their experimental average $\mathrm{D_{eff}}$ whenever the asymmetry factor is $\chi \neq 0$.
The set of coupled LL equations allowed us to simply calculate the eigenfrequencies of the system and determine their differences: $\Delta f$ = $f(+\mathbf k)$ and $f(-\mathbf k)$, which, in turn, are related to the strength of the DMI fields.

\bibliography{main_text}
\bibliographystyle{naturemag}

\section*{Acknowledgements}

This work was supported by the National Science Centre, Poland, Grant No. 2016/23/B/ST3/01430
(SPINORBITRONICS). P.M. acknowledges National Science Centre Grant Beethoven 2 (UMO-2016/23/G/ST3/04196). W.S. acknowledges grant no. 2021/40/Q/ST5/00209 from the National Science Centre, Poland.

\section*{Author contributions statement}

K.G. carried out microstructurization and conducted the electrical conductivity, spin Hall and anomalous Hall experiments. P.O. performed formal analysis and visualization. J.M. is a software developer of modeling software and has performed formal analysis and visualization. P.M. conducted magnetic domain imaging using p-MOKE microscopy measurement and BLS measurements. U.G. conducted BLS measurements. W.S. was responsible for the microstructurization, design, and programming of the measurement methods. T.S. supervised the experimental and theoretical modeling aspects of the project. All authors reviewed the manuscript. 

\section*{Additional information}

\textbf{Competing Interests:} The authors declare that they have no competing interests.

\end{document}